\documentclass[12pt]{article}

\input{tcilatex}

\begin{document}

\begin{center}
\textbf{ON COHERENT OPTICAL PULSE PROPAGATION IN ONE-DIMENSIONAL BRAGG
GRATING }

\bigskip

\vspace{0.5in} Andrei I. Maimistov$^{a}$ \footnote{%
electronic address: amaimistov@hotmail.com}

\vspace{0.2cm} $^{a}$ Department of Solid State Physics, Moscow Engineering
Physics Institute, Kashirskoe sh. 31, Moscow, 115409 Russia
\end{center}

\bigskip

\begin{center}
\textbf{ABSTRACT}
\end{center}

\bigskip

The propagation of the solitary waves in the Bragg grating formed by array
of thin dielectric films is considered. We assume that the thin films of
contain the resonant molecules, which are evolved according to two-level
atoms model, which is used to description of the coherent optical pulses
propagation. There the alternative derivation of the Mantsyzov's equations
is represented.



\newpage\ 

\section{\protect\LARGE Introduction}

In during long time a special kind of gap material has been investigated 
\cite{R1} -\cite{R10}. That was referred to as \emph{resonant Bragg grating} 
\cite{R1}-\cite{R5} or \emph{resonantly absorbing Bragg reflector} (RABR) 
\cite{R6, R7, R8}. In the simplest case RABR consist of the linear
homogeneous dielectric medium containing array of thin films of resonant
atoms or molecules. The distance between neighboring films is $a$, the depth
of film ($l_{f}$) is great less than wave length of radiation spreading in
this structure. The films of two-laves interacted with ultra-short optical
pulse have been considered and investigated in framework of the two-wave
reduced Maxwell-Bloch equations by Mantsyzov at.al. \cite{R1}-\cite{R5}, and
by Kozhekin at.al. \cite{R6, R7, R9}. The existence of the $2\pi $-pulse of
self-induced transparency was demonstrated \cite{R1, R4, R6}. It was found 
\cite{R8} that the bright as well as the dark solitons can exist in the
spectral gap, and the bright solitons can have arbitrary pulse area. If the
two-level atoms density is very high, then the near-dipole-dipole
interaction should be included in the Maxwell-Bloch equations. The effect of
dipole-dipole interaction on the existence of gap solitons in the RABR was
studied in \cite{R10}. The many results of these reviewed in \cite{R9}.
Recently by using numerical simulation the \emph{optical zoomeron} \cite{R14}
was discovered and investigated. Optical zoomeron is similar to soliton,
however it has the velocity which oscillates near some average value.

The two-wave reduced Maxwell-Bloch equations are the base of the
investigation of the solitary wave propagation in the RABR. They have been
deduced by according to coupled-mode theory. There the alternative method of
derivation of the evolutions equations describing the ultrashort optical
pulses propagation in RABR was considered. Hereafter we follow the works 
\cite{R1}-\cite{R8} reasoned that the Bragg resonance corresponds for $%
a=(\lambda /2)m$ where $m=1,2,3,\ldots $. It will be shown that the exact
equations in the form of discrete (recurrent) relations can be obtained. The
number of approximations after that leads to the Mantsyzov's equations. The
two-wave reduced Maxwell-Bloch equations in more general form \cite{R6}-\cite%
{R10} can be deduced by the same procedure also.

\section{\protect\LARGE Transfer-operator approach}

Let us consider the ultra-short optical pulse propagation along the
X-direction of the periodical array of thin films which are replaced in
points $\ldots x_{n-1},x_{n},$ $x_{n+1}\ldots $ (Fig.1). The medium between
films characterized by the dielectric permittivity $\varepsilon $. Hereafter
for the sake of definiteness the TE-wave which has the component of the
electric field parallel to the layers will be considered. All results can be
easily generalized for the case of the TM-polarized waves.

It is suitable the electric and magnetic strengths $\overrightarrow{E},%
\overrightarrow{H}$, and polarization of the two-level atoms ensemble $%
\overrightarrow{P}$ represent in the form of Fourier integrals%
\[
\overrightarrow{E}(x,z,t)=(2\pi )^{-2}\dint\limits_{-\infty }^{\infty }\exp
[-i\omega t+i\beta z]\overrightarrow{E}(x,\beta ,\omega )dtdz, 
\]%
\[
\overrightarrow{H}(x,z,t)=(2\pi )^{-2}\dint\limits_{-\infty }^{\infty }\exp
[-i\omega t+i\beta z]\overrightarrow{H}(x,\beta ,\omega )dtdz, 
\]%
\[
\overrightarrow{P}(x_{n},z,t)=(2\pi )^{-2}\dint\limits_{-\infty }^{\infty
}\exp [-i\omega t+i\beta z]\overrightarrow{P}(x_{n},\beta ,\omega )dtdz. 
\]

Outside off the films the Fourier components of the vectors $\overrightarrow{%
E}(x,\beta ,\omega )$ and $\overrightarrow{H}(x,\beta ,\omega )$\ are
defined by the Maxwell equations. At points $x_{n}$ these values are defined
from the continuity conditions. Thus, the TE-wave propagation can be
considered in framework of the following system%
\begin{equation}
\frac{d^{2}E}{dx^{2}}+\left( k^{2}\varepsilon -\beta ^{2}\right) E=0,
\label{eq1a}
\end{equation}%
\[
H_{x}=-(\beta /k)E,\quad H_{z}=-(i/k)dE/dx,\quad E_{y}=E, 
\]%
with boundary conditions \cite{R11,R12}%
\begin{equation}
E(x_{n}-0)=E(x_{n}+0),\quad H_{z}(x_{n}+0)-H_{z}(x_{n}-0)=4i\pi
kP_{y}(x_{n},\beta ,\omega ),  \label{eq1b}
\end{equation}%
where $k=\omega /c$. The solutions of equation (\ref{eq1a}) in the intervals 
$x_{n}<x<x_{n+1}$ can be written as%
\[
E(x,\beta ,\omega )=A_{n}(\beta ,\omega )\exp [iq(x-x_{n})]+B_{n}(\beta
,\omega )\exp [-iq(x-x_{n})], 
\]%
and%
\[
H_{z}(x,\beta ,\omega )=qk^{-1}\left\{ A_{n}(\beta ,\omega )\exp
[iq(x-x_{n})]-B_{n}(\beta ,\omega )\exp [-iq(x-x_{n})]\right\} , 
\]%
where $q=\sqrt{k^{2}\varepsilon -\beta ^{2}}$ . Hence, the amplitudes $A_{n}$
and $B_{n}$ completely determine the electromagnetic field in RABR. Let us
consider the point $x_{n}$. Electric field at $x=x_{n}-\delta $ ($\delta <<a$%
) is defined by amplitudes $A_{n}^{(L)}$\ and $B_{n}^{(L)}$, the field at $%
x=x_{n}+\delta $ is defined by $A_{n}^{(R)}$\ and $B_{n}^{(R)}$. Continuity
conditions (\ref{eq1a}) result in the following relations between these
amplitudes%
\[
A_{n}^{(R)}\ +B_{n}^{(R)}=A_{n}^{(L)}\ +B_{n}^{(L)}, 
\]%
\[
A_{n}^{(R)}\ -B_{n}^{(R)}=A_{n}^{(L)}\ -B_{n}^{(L)}+4\pi
ik^{2}q^{-1}P_{S,n}, 
\]%
where $P_{S,n}=P_{S}(A_{n}^{(R)}\ +B_{n}^{(R)})$ is the surface polarization
of thin film at point $x_{n}$, which is induced by the electrical field
inside film. Thus one can find%
\begin{equation}
A_{n}^{(R)}\ =A_{n}^{(L)}+2\pi ik^{2}q^{-1}P_{S,n},\quad B_{n}^{(R)}\
=B_{n}^{(L)}-2\pi ik^{2}q^{-1}P_{S,n}.  \label{eq2}
\end{equation}%
With taking account for the strength of electric field outside off the films
we can write%
\begin{equation}
A_{n+1}^{(L)}\ =A_{n}^{(R)}\exp (iqa),\quad B_{n+1}^{(L)}\ =B_{n}^{(R)}\exp
(-iqa).  \label{eq3}
\end{equation}

If the vectors $\psi _{n}^{(L)}=(A_{n}^{(L)},B_{n}^{(L)})$ and $\psi
_{n}^{(R)}=(A_{n}^{(R)},B_{n}^{(R)})$ are introduced, then the relations (%
\ref{eq2}) can be represented in following form%
\[
\psi _{n}^{(R)}=\widehat{U}_{n}\psi _{n}^{(L)},
\]%
where $\widehat{U}_{n}$\ is transfer operator of vector $\psi _{n}^{(L)}$
through film replaced at point $x_{n}$. In general case it is nonlinear
operator. The relations (\ref{eq3}) are represented by vectorial form%
\[
\psi _{n+1}^{(L)}=\widehat{V}_{n}\psi _{n}^{(R)},
\]%
where linear operator of the transferring of the vector $\psi _{n}^{(R)}$
through clearance between adjacent thin films $\widehat{V}_{n}$ is
represented by diagonal matrix%
\[
\widehat{V}_{n}=\left( 
\begin{array}{cc}
\exp (iqa) & 0 \\ 
0 & \exp (-iqa)%
\end{array}%
\right) .
\]%
By this means we define the nonlinear transfer-operator of the vector $\psi
_{n}^{(L)}$ through elementary cell of RABR:%
\begin{equation}
\psi _{n+1}^{(L)}=\widehat{V}_{n}\widehat{U}_{n}\psi _{n}^{(L)}=\widehat{T}%
_{n}\psi _{n}^{(L)}.  \label{eq4}
\end{equation}%
In the case of linear medium transfer-operator is frequently used under
consideration of the one dimensional photonic crystals, in particular, the
distributed feedback structures \cite{R13}.

In (\ref{eq4}) upper index can be omitted, and the equation can be rewritten
in form of the following recurrent relation%
\begin{equation}
A_{n+1}=A_{n}\exp (iqa)+2\pi ik^{2}q^{-1}P_{S,n}\exp (iqa),  \label{eq5a}
\end{equation}%
\begin{equation}
B_{n+1}=B_{n}\exp (-iqa)-2\pi ik^{2}q^{-1}P_{S,n}\exp (-iqa)  \label{eq5b}
\end{equation}%
These recurrent relations are exact equations due to we does not use any
approximation (for example, the slowly varying envelope of electromagnetic
pulses approximation, the long-wave approximation). Furthermore, the surface
polarization of thin film could be calculated on base of deferent suitable
models. Here we will follow the works by B. Mantsyzov at al. \cite{R1}-\cite%
{R5}, and A.Kozhekin, G.Kurizki at.al. \cite{R6}-\cite{R9}, where the
two-levels atom model has been used.

\section{\protect\LARGE \ Linear response approximation }

To demonstrate that the RABR is real gap media it is suitable to obtain the
electromagnetic wave spectrum in linear response approximation. In the
general case we can use the following expression for polarization%
\begin{equation}
P_{S,n}=\chi (\omega )(A_{n}^{(R)}\ +B_{n}^{(R)})  \label{eq6}
\end{equation}%
Substitution of this formula in (\ref{eq5a}),(\ref{eq5b}) leads to%
\begin{equation}
A_{n+1}=(1+i\rho )A_{n}\exp (iqa)+i\rho B_{n}\exp (iqa),  \label{eq7a}
\end{equation}%
\begin{equation}
B_{n+1}=(1-i\rho )B_{n}\exp (-iqa)-i\rho A_{n}\exp (-iqa).  \label{eq7b}
\end{equation}%
Here $\rho =\rho (\omega )=2\pi k^{2}q^{-1}\chi (\omega )=2\pi \omega
c^{-1}\varepsilon ^{1/2}\chi (\omega )$. The wave as collective motion of
the electrical field in grating corresponds to anzats%
\begin{equation}
A_{n}=A\exp (ikna),\qquad B_{n}=B\exp (ikna).  \label{eq8}
\end{equation}%
From (\ref{eq7a}),(\ref{eq7b}) it follows the linear equations are defining
the amplitudes of the wave $A$ and $B$:%
\begin{equation}
A\exp (ika)=(1+i\rho )A\exp (iqa)+i\rho B\exp (iqa),  \label{eq9a}
\end{equation}%
\begin{equation}
B\exp (ika)=(1-i\rho )B\exp (-iqa)-i\rho A\exp (-iqa).  \label{eq9b}
\end{equation}%
Nontrivial solution of this system is exists only if the determinant is not
zero, i.e.,%
\begin{equation}
\det \left( 
\begin{array}{cc}
(1+i\rho )\exp (iqa)-\exp (ika) & i\rho \exp (iqa) \\ 
-i\rho \exp (-iqa) & (1-i\rho )\exp (-iqa)-\exp (ika)%
\end{array}%
\right) =0.  \label{eq10}
\end{equation}

Let us define%
\[
Z=\exp (ika), 
\]%
\[
G=(1+i\rho )\exp (iqa)=(\cos qa-\rho \sin qa)+i(\rho \cos qa+\sin qa) 
\]%
Then the equation (\ref{eq10}) can be rewritten as equation in $Z$:%
\[
Z^{2}-(G+G^{\ast })Z+1=0. 
\]%
The solution of that is 
\[
Z_{\pm }=\func{Re}G\pm i\sqrt{1-(\func{Re}G)^{2}}. 
\]

If $\func{Re}G\leq 1$, then\ $\func{Re}G\pm i\sqrt{1-(\func{Re}G)^{2}}=\cos
ka+i\sin ka$. Hence the wave numbers $k_{\pm }$ are real values and they are
defined from transcendent equation%
\begin{equation}
\cos ka=\cos qa-\rho \sin qa.  \label{eq11}
\end{equation}%
That is the dispersion relation which defines the dependence of wave number $%
k_{\pm }$ versus frequency of harmonic wave propagating in linear RABR.

If $\func{Re}G>1$, then the roots of equation (\ref{eq10}) are real ones. It
means that wave numbers $k_{\pm }$ are imaginary values. The condition $%
\func{Re}G>1$ defines the frequencies of the forbidden zone. The waves with
these frequencies can not propagate in grating. Thus the boundaries of this
forbidden zone follow from the equation%
\begin{equation}
\cos qa-\rho \sin qa=1.  \label{eq12}
\end{equation}

Model of the resonant system containing the thin films defines the explicit
form of the function $\rho =\rho (\omega )$, that results in the different
dispersion relations (\ref{eq11}). It should be pointed out that this
dispersion relation demonstrates a series of gaps in the electromagnetic
wave spectrum.

\section{\protect\LARGE Long-wave and weak nonlinearity approximations }

By using long-wave approximation we can transform the exact equations (\ref%
{eq5a}),( \ref{eq5b}) into the differential equations. To do it is suitable
to introduce the field variables%
\begin{eqnarray*}
A(x) &=&\dsum\limits_{n}A_{n}\delta (x-x_{n}),\quad
B(x)=\dsum\limits_{n}B_{n}\delta (x-x_{n}),\quad \\
P(x) &=&\dsum\limits_{n}P_{S,n}\delta (x-x_{n}),
\end{eqnarray*}%
Using the integral representation for delta-function%
\[
\delta (x)=(2\pi )^{-1}\dint\limits_{-\infty }^{\infty }\exp (ikx)dk, 
\]%
one can obtain the following expression 
\begin{eqnarray*}
A(x) &=&(2\pi )^{-1}\sum_{n}A_{n}\dint\limits_{-\infty }^{\infty }\exp
[ik(x-x_{n})]dk= \\
&=&(2\pi )^{-1}\dint\limits_{-\infty }^{\infty }\exp (ikx)\sum_{n}A_{n}\exp
(-ikx_{n})dk.
\end{eqnarray*}%
From that 
\[
A(k)=\sum_{n}A_{n}\exp (-ikx_{n})=\sum_{n}A_{n}\exp (-ikan), 
\]%
and by the same way one get%
\[
B(k)=\sum_{n}B_{n}\exp (-ikan),\quad P(k)=\sum_{n}P_{S,n}\exp (-ikan). 
\]

The definition of these spatial Fourier components results in the
periodicity conditions, for example,%
\begin{eqnarray}
A(k) &=&\sum_{n}A_{n}\exp (-ikan)\exp (\pm 2\pi in)=  \nonumber \\
&=&\sum_{n}A_{n}\exp (-ikan\pm 2\pi in)=  \label{eq14} \\
&=&\sum_{n}A_{n}\exp [-ian(k\pm 2\pi /a)]=A(k\pm 2\pi /a).  \nonumber
\end{eqnarray}%
From recurrent equations (\ref{eq5a}),(\ref{eq5b}) we have%
\[
A(k)\exp (ika)=A(k)\exp (iqa)+i\kappa P(k)\exp (iqa),
\]%
\[
B(k)\exp (ika)=B(k)\exp (-iqa)-i\kappa P(k)\exp (-iqa),
\]%
or%
\begin{equation}
\left[ \exp \{ia(k-q)\}-1\right] A(k)=i\kappa P(k),  \label{eq15a}
\end{equation}%
\begin{equation}
\left[ \exp \{ia(k+q)\}-1\right] B(k)=-i\kappa P(k).  \label{eq15b}
\end{equation}

The linear approximation for polarization $P(k)=\chi (\omega )[A(k)+B(k)]$\
leads us to linear dispersion law (\ref{eq10}) once again. The system of
equations (\ref{eq15a}),(\ref{eq15b}) is equivalent of discrete equations (%
\ref{eq5a}),(\ref{eq5b}). At this point we have not approximation, apart
from assumption of thin films width.

If the thin film array was absent then the dispersion relation would be%
\[
\cos ka=\cos qa. 
\]%
From that, for right wave with amplitude $A(k)$ we have $k=q$, whereas for
wave with amplitude $B(k)$ propagation in opposite direction we have $k=-q$.
Let us suppose that the polarization of the thin film array produce the
little change in wave vectors, i.e., for right wave it is $k=q+\delta k$ and
for the opposite wave it is $k=-q+\delta k$. Let us chouse the value of $q$
near one of the Bragg resonances, to say, $q=2\pi /a+\delta q$, where $%
\delta q\ll 2\pi /a$. In this case the equations (\ref{eq5a}),(\ref{eq5b})
take the form%
\[
\left[ \exp (ia\delta k)-1\right] A(q+\delta k)=i\kappa P(q+\delta k), 
\]%
\[
\left[ \exp (ia\delta k)-1\right] B(-q+\delta k)=-i\kappa P(-q+\delta k). 
\]%
By using the periodicity conditions (\ref{eq14}) these equations are
rewritable as%
\[
\left[ \exp (ia\delta k)-1\right] A(\delta q+\delta k)=i\kappa P(\delta
q+\delta k), 
\]%
\[
\left[ \exp (ia\delta k)-1\right] B(-\delta q+\delta k)=-i\kappa P(-\delta
q+\delta k). 
\]%
After changing of the variables $\delta k=\delta \tilde{k}\pm \delta q$ we
have%
\begin{equation}
\left[ \exp \{ia(\delta \tilde{k}-\delta q)\}-1\right] A(\delta \tilde{k}%
)=i\kappa P(\delta \tilde{k}),  \label{eq16a}
\end{equation}%
\begin{equation}
\left[ \exp \{ia(\delta \tilde{k}+\delta q)\}-1\right] B(\delta \tilde{k}%
)=-i\kappa P(\delta \tilde{k}).  \label{eq16b}
\end{equation}%
The long-wave approximation means that nonzero values of the spatial Fourier
amplitudes are located near zero value of argument. Hence, we can assume $%
a\delta \tilde{k}\ll 1$ in exponential functions, that results in the
following approximate equations%
\begin{equation}
ia(\delta \tilde{k}-\delta q)A(\delta \tilde{k})=i\kappa P(\delta \tilde{k}),
\label{eq17a}
\end{equation}%
\begin{equation}
ia(\delta \tilde{k}+\delta q)B(\delta \tilde{k})=-i\kappa P(\delta \tilde{k}%
).  \label{eq17b}
\end{equation}%
If now we return in to spatial variable, then equations (\ref{eq17a}),(\ref%
{eq17b}) lead us to the equations of coupled wave theory%
\begin{equation}
\partial A/\partial x=i\delta qA(x)+i\kappa a^{-1}P(x),  \label{eq18a}
\end{equation}%
\begin{equation}
\partial B/\partial x=-i\delta qB(x)-i\kappa a^{-1}P(x).  \label{eq18b}
\end{equation}

In these equations the both fields $A(x),B(x),P(x)$ and parameters $\delta q$%
, $\kappa $ are the functions of frequency $\omega $. To obtain final system
of equations in the coordinate and time variables the inverse Fourier
transformation would be done. However, frequently the assumption of a slowly
varying it time scale envelopes of electromagnetic pulses and polarization
used \cite{R14}. This approximation allows simplifying the system of coupled
wave equations under consideration.

\section{\protect\LARGE Slowly varying envelopes approximations }

Slowly varying envelopes approximation minds that we have to deal with
narrow wave packets or, what is the same, with quasiharmonic waves \cite{R14}%
. For example, for the quasiharmonic wave the electric field is represented
by expression%
\[
E(x,t)=\mathcal{E}(x,t)\exp [-i\omega _{0}t],
\]%
where $\omega _{0}$ is carrier wave frequency. The electric field $E(x,t)$
and Fourier components of the envelope of the pulse $\tilde{E}(x,t)$ are
related by the following relationship%
\begin{eqnarray*}
E(x,t) &=&(2\pi )^{-1}\dint\limits_{-\infty }^{\infty }E(x,\omega )\exp
(-i\omega t)d\omega = \\
&=&(2\pi )^{-1}\dint\limits_{-\infty }^{\infty }\mathcal{E}(x,\omega )\exp
[-i(\omega +\omega _{0})t]d\omega = \\
&=&(2\pi )^{-1}\dint\limits_{-\infty }^{\infty }\mathcal{E}(x,\omega -\omega
_{0})\exp (-i\omega t)d\omega ,
\end{eqnarray*}%
where the function $E(x,\omega )$ is nonzero one if $\omega $ belongs to
interval $(\omega _{0}-\Delta \omega ,\omega _{0}+\Delta \omega )$ with $%
\Delta \omega \ll \omega _{0}$.\ It leads to $E(x,\omega +\omega _{0})=%
\mathcal{E}(x,\omega )$. Thus, if we have some relation for $E(x,\omega )$,
then in order to carry out the needed relation for\ $\mathcal{E}(x,\omega )$
to be done shifting $\omega $\ $\rightarrow \omega _{0}+\omega $ in all
functions of $\omega $.

Let be%
\[
A(x,t)=\mathcal{A}(x,t)\exp (-i\omega _{0}t),\quad B(x,t)=\mathcal{B}%
(x,t)\exp (-i\omega _{0}t), 
\]%
\[
P(x,t)=\mathcal{P}(x,t)\exp (-i\omega _{0}t). 
\]%
For the Fourier components \ $\mathcal{A}(x,\omega ),\mathcal{B}(x,\omega )$
and $\mathcal{P}(x,\omega )$\ from (\ref{eq18a}),(\ref{eq18b}) it follows%
\begin{equation}
\partial \mathcal{A}(x,\omega )/\partial x=i\delta q(\omega _{0}+\omega )%
\mathcal{A}(x,\omega )+i\kappa (\omega _{0}+\omega )a^{-1}\mathcal{P}%
(x,\omega ),  \label{eq19a}
\end{equation}%
\begin{equation}
\partial \mathcal{B}(x,\omega )/\partial x=-i\delta q(\omega _{0}+\omega )%
\mathcal{B}(x,\omega )-i\kappa (\omega _{0}+\omega )a^{-1}\mathcal{P}%
(x,\omega )  \label{eq19b}
\end{equation}%
Inasmuch as in this expression $\mathcal{A}$, $\mathcal{B}$ and $\mathcal{P}$%
\ are distinguished from zero at $\omega \ll \omega _{0}$, one can use the
expansions:%
\begin{equation}
\delta q(\omega _{0}+\omega )\approx q_{0}-2\pi /a+q_{1}\omega +q_{2}\omega
^{2}/2,\quad \kappa (\omega _{0}+\omega )a^{-1}\approx K_{0}.  \label{eq20}
\end{equation}%
where $q_{n}=d^{n}q/d\omega ^{n}$ at $\omega =\omega _{0}$. In particular,
the $q_{1}^{-1}=v_{g}$ is the group velocity, $q_{2}$ takes account the
group-velocity dispersion.

With taking expansions (\ref{eq20}) into account we can write following
equations which describe the evolution of slowly varying envelopes%
\begin{equation}
i\left( \frac{\partial }{\partial x}+\frac{1}{v_{g}}\frac{\partial }{%
\partial t}\right) \mathcal{A}-\frac{q_{2}}{2}\frac{\partial ^{2}}{\partial
t^{2}}\mathcal{A}+\Delta q_{0}\mathcal{A}=-K_{0}\mathcal{P},  \label{eq21a}
\end{equation}%
\begin{equation}
i\left( \frac{\partial }{\partial x}-\frac{1}{v_{g}}\frac{\partial }{%
\partial t}\right) \mathcal{B}+\frac{q_{2}}{2}\frac{\partial ^{2}}{\partial
t^{2}}\mathcal{B}-\Delta q_{0}\mathcal{B}=+K_{0}\mathcal{P},  \label{eq21b}
\end{equation}%
where $\Delta q_{0}=q_{0}-2\pi /a$. To do next step it need to chouse the
model for thin film medium. It can be enharmonic oscillators, the two- or
three-levels atoms, the excitons of molecular chains, nano-particles, the
quantum dots, and so on. Here we consider the two-level atom model.

\section{\protect\LARGE Two level atoms approximations }

The two-level atom state defines by the density matrix $\hat{\rho}$. The
matrix element $\rho _{12}$ describes the transition between the ground
state $\left\vert 2\right\rangle $ and excited state $\left\vert
1\right\rangle $ , $\rho _{22}$ and $\rho _{11}$ represents the population
of theses states. Evolution of the two-level atom is governed by the Bloch
equations \cite{R15}.%
\begin{equation}
i\hbar \frac{\partial }{\partial t}\rho _{12}=\hbar \Delta \omega \rho
_{12}-d_{12}(\rho _{22}-\rho _{11})A_{in},  \label{eq22a}
\end{equation}%
\begin{equation}
i\hbar \frac{\partial }{\partial t}(\rho _{22}-\rho _{11})=2\left(
d_{12}\rho _{21}A_{in}-d_{21}\rho _{12}A_{in}^{\ast }\right) .  \label{eq22b}
\end{equation}
In these equations $A_{in}$ is the electric field interacting with two-level
atom. In the problem under consideration we have $A_{in}=\mathcal{A}+%
\mathcal{B}$\ and%
\[
K_{0}\mathcal{P}=\frac{2\pi \omega _{0}n_{at}d_{12}}{cn(\omega _{0})}%
\left\langle \rho _{12}\right\rangle . 
\]%
There the cornerstone brackets denote summation over all atoms within a
frequency detuning $\Delta \omega $ from center of the inhomogeneously
broadening line, $n(\omega _{0})$ is the refractive index of the medium
containing the array of thin films, $n_{at}$ is effective density of the
resonant atoms in films. This value is defined thought bulk density of atoms 
$N_{at}$, film width $l_{f}$ and lattice spacing by formula $%
n_{at}=N_{at}(l_{f}/a)$.

Let us suppose the group-velocity dispersion is of no importance. The
resulting equations are the \emph{two-wave reduced Maxwell-Bloch equations}.
It is suitable to introduce the normalized variables%
\[
e_{1}=t_{0}d_{12}\mathcal{A}/\hbar ,\quad e_{2}=t_{0}d_{12}\mathcal{B}/\hbar
,\quad x=\zeta v_{g}t_{0},\quad \tau =t/t_{0}. 
\]%
The normalized two-wave reduced Maxwell-Bloch equations take the following
form%
\begin{equation}
i\left( \frac{\partial }{\partial \zeta }+\frac{\partial }{\partial \tau }%
\right) e_{1}+\delta e_{1}=-\gamma \left\langle \rho _{12}\right\rangle ,
\label{eq23a}
\end{equation}%
\begin{equation}
i\left( \frac{\partial }{\partial \zeta }-\frac{\partial }{\partial \tau }%
\right) e_{2}-\delta e_{2}=+\gamma \left\langle \rho _{12}\right\rangle ,
\label{eq23b}
\end{equation}%
\begin{equation}
i\frac{\partial }{\partial \tau }\rho _{12}=\Delta \rho _{12}-ne_{in},
\label{eq23c}
\end{equation}%
\begin{equation}
\frac{\partial }{\partial \tau }n=-4\func{Im}(\rho _{12}e_{in}^{\ast }),
\label{eq23d}
\end{equation}%
where $\gamma =t_{0}v_{g}/L_{a}$, $\delta =t_{0}v_{g}\Delta q_{0}$, $%
L_{a}=(cn(\omega _{0})\hbar )/(2\pi \omega _{0}t_{0}n_{at}|d_{12}|^{2})$\ is
resonant absorption length and $\Delta =\Delta \omega t_{0}$\ is normalized
frequency detuning. We use the following variables $n=\rho _{22}-\rho _{11}$%
, $e_{in}=e_{1}+e_{2}$.

Let define new variables according to following expressions%
\[
e_{in}=e_{1}+e_{2}=f_{s}\exp (i\delta \tau ),\quad e_{1}-e_{2}=f_{a}\exp
(i\delta \tau ),\quad \rho _{12}=r\exp (i\delta \tau ).
\]%
The system of equations (\ref{eq23a}) - (\ref{eq23d}), can be rewritten as%
\begin{equation}
\frac{\partial f_{s}}{\partial \zeta }+\frac{\partial f_{a}}{\partial \tau }%
=0,  \label{eq25a}
\end{equation}%
\begin{equation}
\frac{\partial f_{a}}{\partial \zeta }+\frac{\partial f_{s}}{\partial \tau }%
=2i\gamma \left\langle r\right\rangle ,  \label{eq25b}
\end{equation}%
\begin{equation}
i\frac{\partial }{\partial \tau }r=(\Delta +\delta )r-nf_{s},  \label{eq25c}
\end{equation}%
\begin{equation}
\frac{\partial }{\partial \tau }n=-4\func{Im}(rf_{s}^{\ast }).  \label{eq25d}
\end{equation}%
From (\ref{eq25a}) it follows%
\[
\frac{\partial f_{a}}{\partial \zeta }=-\frac{\partial f_{s}}{\partial \tau }%
,
\]%
It allows rewrite the system (\ref{eq25a})-(\ref{eq25d}) in another form:%
\begin{equation}
\frac{\partial ^{2}f_{s}}{\partial \zeta ^{2}}-\frac{\partial ^{2}f_{s}}{%
\partial \tau ^{2}}=-2i\gamma \left\langle \frac{\partial r}{\partial \tau }%
\right\rangle ,  \label{eq26a}
\end{equation}%
\begin{equation}
i\frac{\partial }{\partial \tau }r=(\Delta +\delta )r-nf_{s},  \label{q26b}
\end{equation}%
\begin{equation}
\frac{\partial }{\partial \tau }n=-4\func{Im}(rf_{s}^{\ast })  \label{eq26c}
\end{equation}

If we assume that the inhomogeneous broadening is absent, i.e., the
hypothesis of a sharp atomic resonant transition is true, this system is
reduced to Sine-Gordon equation $\delta +\Delta =0$ \cite{R1}. In \cite{R4}
the steady state solution of (\ref{eq26a})-(\ref{eq26c}) with taking
inhomogeneous broadening into account was found.

\section{\protect\LARGE Conclusion}

There the derivation of the equations which are describing the ultrashort
pulses propagation in one dimensional resonant Bragg grating (or RABR) was
considered. The grating consists in thin films array embedded into linear
dialectical medium. The exact discrete equation for amplitudes electric
field inside films was obtained. The long-wave approximation and slowly
varying envelope of the pulses approximation result in the system of the
coupled wave equations. At this point it need to chouse the model for thin
film medium. In \cite{R1}- \cite{R10} the two-level atom model has been
used. However, other models could be considered that results in new kind of
RABRs. For example, array of thin films of the resonant optical materials
with embedded metal nanostructures \cite{R16}. would be described by the
following system%
\begin{equation}
i\left( \frac{\partial }{\partial \zeta }+\frac{\partial }{\partial \tau }%
\right) e_{1}+\delta e_{1}=-\gamma \left\langle \sigma \right\rangle ,
\label{eq27a}
\end{equation}%
\begin{equation}
i\left( \frac{\partial }{\partial \zeta }-\frac{\partial }{\partial \tau }%
\right) e_{2}-\delta e_{2}=+\gamma \left\langle \sigma \right\rangle ,
\label{eq27b}
\end{equation}%
\begin{equation}
i\frac{\partial }{\partial \tau }\sigma -\Delta \sigma +\mu |\sigma
|^{2}\sigma =(e_{1}+e_{2}).  \label{eq27c}
\end{equation}%
where $\mu $ is coupling constant associated with enharmonicity of the
plasmonic oscillations. Second example of the resonant model corresponds
with three-level atom and two-frequency electromagnetic field. Resulting
system of the basic equations is similar to system considered in \cite{R17},
where the propagation of the polarized waves in RABR has been studied. A
fascinating model was proposed by A. Zabolotskii \cite{R18}, where thin
films are replaced with J-aggregates. Array of thin ferroelectric films
represent the gap-medium where the switching wave propagation could be
expected.

It should be noted that in \cite{R6} - \cite{R9} the two-wave reduced
Maxwell-Bloch equations contain extra terms taking account the dispersion of
polarization. These terms can be provided by this analysis also.

\section*{Acknowledgment}

I would like to express my gratitude to Dr. B.I. Mantsyzov, Dr. A.A.
Zabolotskii, Prof. I.R. Gabitov, and Prof. J-G. Caputo for enlightening
discussions. I am grateful to the \textit{Laboratoire de Math\'{e}matiques,
INSA de Rouen} for hospitality and support.

\newpage

\section*{FIGURE CAPTIONS}

Fig. 1. Model of periodical nonlinear medium corresponding with resonantly
absorbing Bragg reflector.

\end{document}